\documentclass[11pt,showkeys,a4paper]{revtex4-2}

\usepackage{bm}
\usepackage[colorlinks=true, citecolor=blue, urlcolor=blue]{hyperref}
\usepackage{epsf,graphics,graphicx}
\usepackage{amsmath}
\usepackage{latexsym}
\usepackage{amssymb,amsmath,graphicx}
\usepackage{amsfonts}
\usepackage[caption=false]{subfig}
\usepackage{epsfig,latexsym,graphicx}
\usepackage{amssymb,color}
\newtheorem{remark}{Remark}

\begin{document}
\title{\Large  %On the existence of Scaling laws across Indian districts:\\ A new prospect for urban scaling \\ 
Multi-scale analysis of  rural  and urban areas: A case study of Indian districts}

\author{Abhik Ghosh}
\email{abhik.ghosh@isical.ac.in}
\affiliation{Indian Statistical Institute, Kolkata 700108, India}

\author{Souvik Chattopadhay}
%\email{souvikchatterjee812@gmail.com}
\affiliation{Indian Statistical Institute, Kolkata 700108, India}

\author{Banasri Basu}
\email{sribbasu1@gmail.com}
\affiliation{Indian Statistical Institute, Kolkata 700108,  India}

\date{\today}

\begin{abstract}
\noindent\textbf{Abstract:} 	
\textcolor{black}{
%Urban scaling analysis in generally performed based on cities. 
%In this study, we have  explored a scaling analysis based  % new prospect for u scaling analysis based 
%on relatively larger local administrative units 
%which are independently functional, within a country.
%Urbanization  is the population shift from rural to urban areas, the corresponding decrease in the proportion of people living in rural areas, and the ways in which societies adapt to this change. %It is an index of transformation from traditional rural economies to modern
%industrial one.
\noindent 
It is well known that  the urban systems, in particular cities,  display scaling behaviour regarding  socio-economic, infrastructural  and individual basic services indicators.  However, understanding urbanisation and the links between rural and urban areas is fundamental to making the most of the global transformations happening around the world.
In this context, it is important to study the  scaling laws based on both the urban and rural regions, going beyond cities.   This paper explores the extension of the idea of allometric urban scaling law to study  the scaling  behaviour of  Indian districts, with both the urban and rural population.  To proceed, we have chosen  districts ( both rural and urban)  of India, a relatively larger local administrative units, which are more or less independently functional within a country. 
This interdisciplinary  work focus on  the scaling analysis of various socio-economic indicators (SEIs) corresponding to the size ( population) of  four distinct urbanization classes, namely rural, semi-rural, semi-urban and urban districts.
The scaling exponents  ($\beta$ ) were estimated for each classes for the years 2001 and 2011 along with their goodness-of-fit measured by the $R^2$ values. 
Our rigorous statistical analysis indicates  that the scaling laws indeed exist even at the district level for most of the SEIs considered, 
related to education, employment, housing, health, etc.;  the $R^2$ values obtained for these SEIs are very high (often greater than 0.8 or 0.9)  in both the the years. 
Moreover,  linearity of the scaling factors   have  been statistically tested  and it has been found, at 95\%  level of confidence, 
that  not all the SEIs behave linearly ($\beta=1$) ; some of them are characterized  by super-linear ($\beta >1$) 
behaviour and some behave sub-linearly ($\beta < 1$) .  
Statistical hypothesis tests have also been performed to test the equality of two scaling factors corresponding to 
two distinct classes and two different years to understand the differences in scaling relationships among increasing urbanisation classes
and their changes over time}. 
%\vspace{3cm}
\end{abstract}

\maketitle

%\bigskip\bigskip \noindent
%{\bf{Acknowledgment}}:\\

%This  research is supported by a research grant (No. CRG/2019/001461) from the Science and Engineering Research Board (SERB), Government of India, India.  
%SC thanks Science and Engineering Research Board (SERB), Government of India for financial support through Junior Research Fellowship (Grant No.CRG/2019/001461). 
%\end{document}

\newpage
\section{Introduction}

\noindent 
Scaling is one of the simplest rules that describes the underlying behaviour of a complex system \cite{2017rule}. 
Scaling, as a manifestation of the fundamental dynamics,  has been conducive to % in helping 
scientists gather deeper insights about  problems 
ranging across the entire spectrum of science and technology. 
\textcolor{black}{The analytic framework of scaling may help one to model the dependence of aggregate properties of systems on their sizes.}
In the literature we can find numerous examples of scale invariance properties including earthquakes, clouds, networks, etc. \cite{bak2002,siebesma2000,dorogovstev2001}.  
Scale invariance seems to be widespread in natural systems. 
%Scale invariance  has been found to hold empirically in  a number of complex systems, and the 
Correct evaluation of the scaling exponents is of primary importance in assessing  the existence of universality classes \cite{stanley2000}. 
It has been widely used in biological, ecology and physical complex systems \cite{west_gb2017}. 

%The origin of allometric scaling law ($Y=Y_0N^{\beta}$ for a variable $Y$ in a population of size $N$) in biology is well explained \cite{west_gb2005}; using this biological metaphor in urban scaling; one can successfully predict many underlying features like economic diversity, innovation, productivity  \cite{betten_innov2007} and many more. 
%However, unlike biological scaling that follows sub-linear ($\beta<1$) relation, 
%scaling exponent $\beta$ of urban scaling takes sub-linear  ($\beta <1$),   super-linear ($\beta>1$) and linear ($\beta=1$) values. 

\noindent
\textcolor{black}{
Scaling has also been used to model urban growth in cities through self-similarity \cite{Batty/etc:1989,Batty/Longley:1994,Frankhauser:2008,batty_2008, chen2006, betten2013b,  betten2017b, betten2012stat, betten_2010_unified, betten_2016_europe};  
the  essential properties of cities in terms of their infrastructure and socio-economics are functions of their population size in a scale invariant way
and these scale transformations are common to all urban systems.
Urban scaling shows how urban indicators change with the size of city (generally population) 
and predicts several underlying features  like economic diversity, innovation, productivity  and many more \cite{betten_innov2007, settle_gang} .}
It is important to study such scaling  laws  of various socio-economic  indicators  associated with the urbanisation in a country.

%\end{document}
\noindent 
Urbanisation is a socio-economic transformation that converts rural settlements to urban settlements 
by building  infrastructural facilities, increasing  the population of urban areas in comparison to rural areas. 
\textcolor{black}{
Beyond such simplistic understanding, it is indeed an economic, physical, socio-cultural change 
and  economic activities change giving way to agglomeration and subsequent urbanization. 
Even after cities emerged, however, a large majority of people lived and worked in rural areas. }
%It was not until large-scale industrialization began in the eighteenth century that cities really began to boom.} 
%Nearly half of all people now live in urban area. 
%
%The world has urbanized rapidly since 1950. % and projections indicate that it will continue to urbanize in
%the coming decades. 
%In 1950 the world was mostly rural: more than two-thirds of people lived in rural areas %settlements
%and less than one-third in urban settlements. 
%\textcolor{black}{Later  in 2014,  it has been found that over half of the global population has become urban. However, this distribution is expected to shift further towards urban
%areas over the next 35 years so that, 
It has been projected that by 2050, the world’s population will be one-third rural and two-thirds urban,
roughly the reverse of the situation in the mid-twentieth century \cite{un2014}.  %Northern America (82\%), Latin America and the Caribbean (81\%), Europe (74\%) are the most urbanized regions
%whereas Asia (50\%) and Africa (43\%) have low degree of urbanisation \cite{un2018}}. 

\noindent  
It is interesting to note that Asia and Africa together cover 90\% \cite{un2018} of world's rural population with the largest rural populated country being India (893 million). 
In recent years, however, Asian and African countries are portraying  a rapid accelerating urbanisation. 
For example, the population residing in urban areas in India was 11.4\% in 1901, which increased to 28.53\% by the 2001 census \cite{india_pop}. % [xx ref?? xx]. 
In the year 2011,  for the first time since independence, the absolute $increase$ in population in India was more in urban areas than  in rural areas.  
Level of urbanization increased from 27.81 \% in 2001 Census to 31.16 \% in 2011 Census \cite{india_pop}. 
This  data inspired us to study a comparative scaling analysis of the distributional  features of various socio-economic indicators associated with  the urban and rural population in India, a country  
with diverse geographical, economic and cultural  background \cite{ro_districts}. 
\textcolor{black}{
The effect of urbanization can be realized by the  proper information of socio-economic trends 
which can, in turn, help us studying its sustainability over a longer  period of time.
To be specific, understanding the scaling features of various socio-economic indicators, associated with  both the urban and rural population in a country, is essential to make sure that the appropriate resources 
and services are available where they are needed. This challenging study  may provide a new prospect of urban scaling analysis.}
%To be specific, understanding the distribution of people in rural and urban areas %settlements 
%in a given country,  associated with various socio-economic indicators, is essential to make sure that the appropriate resources 
%and services are available where they are needed.
%}

\noindent 
{\color{black}{It is well known that in the urban scaling analysis, scaling laws are mostly studied in respect of cities, where  cities are  used as independent functioning population units with the underlying rationale being the idea of agglomeration.
%Scaling laws are mostly studied in respect of $cities$ and we know that Cities are mostly used as the independent functioning population units in urban scaling analysis,
%with the underlying rationale being the idea of ``agglomeration".
Despite advances in the study of science in cities, we still lack a unified theoretical framework for the spatial distributions of various elements within cities and macroscopic scaling laws across cities. 
These results  indicate  the existence of high variability of scaling exponents with respect to variations in the definition of city or urban areas
\cite{Cottineau/etc:2017,Louf/Barthelemy:2014}.  
There is another  practical difficulty in studying the socio-economic properties of cities in respect of scaling analysis,
since most official statistics on socio-economic indicators pertain to somewhat arbitrarily defined administrative units, which, at some level, are not cities at all. 
Examples include counties or census tracts in the USA, several forms of local authority in the UK, prefectures in Japan, 
prefecture and districts in China, districts in India and municipalities in many European and South American nations \cite{Bettencourt/etc:2013}.
}}		

\noindent  
With this backdrop one may  try to formulate a unified framework  of $scaling$,  comprising both the rural and urban population, implementing  the idea of $city$ $based$ urban scaling analysis.
%This makes one understand  the significance of extending the idea of urban scaling analysis  to the rural case. With this backdrop one must try formulate a unified framework   of scaling analysis comprising both the rural and urban population.  
It is important to perceive how  the change in   various rural- and urban- level  factors  with respect to the population-size  can be instrumental in  creation of a unified theory of scaling  in the rural and urban areas. This inspired us to extend the idea of $city$ $based$ urban scaling analysis at the level of Indian districts consisting  of both the urban and rural population.
These studies may reveal the general characteristics of infrastructure and services in the urban as well as  rural areas of a nation like India.
%In this era  of rapid urbanization, in a country  where many people  still live in rural areas, the  study of urban scaling analysis based on  cities only  is not enough and motivates one to study the scaling laws from a  different perspective.  
% It  is worthy to formulate a unified framework where  the scaling relations can be studied in respect of both the rural and urban population simultaneously.  
There are a few studies on urban scaling in this direction.  Recently urban scaling has been studied in Indian slums using Census data \cite{world_dev}, %\textcolor{black}{From a different perspective (other than cities), 
the scaling law has  been studied  in intra-urban administrative units \cite{china2020} and also  there is a scaling analysis of rural housing land transition under China's
rapid urbanization \cite{habitat} . 
However, it is unclear whether scaling law exists for a more general administrative unit which includes both rural and urban areas. 
In general, for the  administrative purpose and the ease of working in a local manner, 
the states or provinces (first tier division i)  in a large country, like China or India, 
are subdivided  into more smaller (second tier) strata called counties or districts. 
Although there is a wide heterogeneity among the administrative units (states or districts) in a country,  the
 local administrative units can be seen as socially constructed strata  \cite{ro_districts}, which serve as spatial scenarios for social and economic processes \cite{lopez2008}. 
Such administrative divisions are often regions of a country that are granted a certain degree of autonomy and manage themselves through local governments.
So, the  scaling analysis  done at the level of such local administrative units in a country (e.g., districts in India)  is expected to be more comprehensive as well as challenging. 

\noindent
{\color{black}{
In this paper, our main goal is to empirically verify if the scaling law  exists across somewhat different and larger administrative units (other than cities). 
For this purpose, we consider all the Indian districts, as administrative functioning units, which are surely independent, 
and investigate the existence of the scaling laws of a few important socio-economic variables in respect to their population. However, the geographic, cultural, economic or political diversity of Indian districts are well known; 
most districts in India have both rural and urban populations, except for a few large cities which are districts by themselves.
And,}} it has recently been reported  that consideration of all the districts, of all the states, in India do not show agglomeration economies 
as its geographical diversity is critical to emergence of  the scaling law \cite{betten_2019_geo}. 
\textcolor{black}{So, we appropriately classify the districts in more homogeneous classes with respect to their urbanisation status.}
Different classification criteria \cite{nelson_class1955} have been proposed and widely applied  
to find the distinct types of urban, suburban regions of USA and China \cite{mikelb_class2004,liux_appl2020}.  
%Such a classification allows us to study how an element of a class,  behaves  in a similar way or differently with other elements of the same or different class. 
In the context of urbanisation in India, features of classification may serve as an important tool  in studying the diverse nature of Indian districts. 
Economic, demographic, and spatial structures may play an important role to find out the distinct classes.  
As our  primary  focus is to see the effect of urbanisation in Indian states,  
we set an urbanisation index,  the ratio of urban and rural population of each district in India, 
as the  classification parameter to find out four distinct classes,
\textcolor{black}{namely the rural, semi-rural, semi-urban and urban (see Section \ref{SEC:classification} for details).
In the context of  rapid urbanisation in India,  it is very important and also challenging to study the scaling laws in respect to 
these four urbanization classes separately. The result of linear or non-linear scaling (super-linear or sub-linear) 
may help one to frame future unified policies for different districts in India.}

\noindent 
%Accordingly, the objective  of the present paper is mainly two fold. 
{\color{black}{In summary, the contributions  of the present paper is %thus 
many fold.}}
\begin{itemize}
	\item Firstly,  we statistically test for the existence of scaling laws at district level administrative units of India, 
	\textcolor{black}{comprising both the rural and urban population,} 
	by using the census data of the years 2011 and 2001 for various socio-economic indicators. 
As far as  our knowledge goes, in the relevant literature we find no study %there is no study in this context we In t has never been done in the literature 
with such a large amount of data in Indian context. 
\textcolor{black}{Also the existing studies were mostly focused on cities whereas 
	we empirically demonstrate  the existence of scaling laws, in particular allometric scaling laws, across the districts of India within appropriate urbanisation classes,
	suggesting a multi-scale analysis in the spatial morphology leading to  a new prospect for urban scaling analyses. }

\item \textcolor{black}{
Secondly, we estimate the scaling exponent $\beta$ for all the distinct urbanization classes.   We adopt the method of statistical hypothesis testing to ensure  the robustness of our. analysis in estimating  the power law exponent $\beta$. %, we perform the statistical hypothesis testing. 
%Secondly, we perform statistical hypothesis testing to study the power law exponent $\beta$. %agglomeration effect.  
%In this paper,  for the robustness of our analysis, 
We statistically test for the linearity ($\beta=1$) against sub or super linearity ($\beta< 1$  or $\beta >1$)  of the scaling index 
rather than deciding just based on their values. It has already been shown that the process of measuring the power law exponent needs 
a rigorous statistical approach  \cite{Leitao/etc:2016}.}
It helps us to put confidence in our conclusion taking care of the sampling fluctuations and other sorts of distortion present in the empirical data.

\item Finally, %We also compare the scaling indices  in two different urbanisation classes in two different time points 
%by appropriate statistical tests to deduce more reliable inferences. 
\textcolor{black}{based on these results, we comment, with statistical confidence,} 
on the allometric scaling laws of  distinct  urbanisation  classes in India. Our scaling analysis portray  the effects of  urbanisation  on  different socio-economic parameters along with their changes from the year 2001 to 2011.  
\end{itemize}

\noindent 
 The rest of the paper is organized as follows. 
 Section \ref{SEC:matrl} is devoted to explain the data and the methodology that has been used for this study.
 Section \ref{SEC:reslt} presents detailed discussion on the  results of the analyses. 
 Finally, we summarize and conclude the paper in \ref{SEC:conclusion}.

%\textcolor{red}{The rest of the paper organised as: Section\ref{matrl} for research data, methods and section \ref{reslt} for result and discussion}
%\newpage
\section{Materials and Methods}\label{SEC:matrl}

\subsection{Research Data}
\label{SEC:Data}
\noindent
We consider the data from Indian census which is conducted in every 10 years to capture a detailed picture of 
demographic, economic and social conditions of all persons in the country pertaining to that specific time. 
The raw census data for the years 2011 and 2001 are obtained from the \textit{Primary Census data and Digital library} (www.censusindia.gov.in). 
Following the stratified administrative structure of India, we use the districts as the smaller second-tier units within each state (first-tier strata) of India. 
According to the final census in 2011, there were  29 states and 7 Union territories (UT)  in India, which consist of the pool of strata in our analysis. 
It is important to note that some states, created at a later time, were not present in the earlier round of census (at the year 2001). 
The total number of districts comprising  all the states (and UTs) of India was 640 and 593 in the years 2011 and 2001, respectively.  

\noindent 
The data shows wide inter-State disparity in urbanization level \cite{hand_india}.
In terms of overall urban population, Maharashtra had the largest urban population of 50.8 million followed by Uttar Pradesh, 
which had an urban population of 44.5 million in 2011. If we look at the level of urbanization, defined as urban population as a proportion of total population, 
Goa was the most urbanized state with 62.17\% urbanization in 2011 followed by Mizoram at 52.11\% urban population. 
Among the Union Territories, Delhi had urbanization level of 97.50\%  followed by Chandigarh with an urbanization level of 97.25\% in 2011. 
Himachal Pradesh had the lowest urbanization with only 10.03\% population living in urban areas in 2011, followed by Bihar (11.29\%).
 
 \noindent 
According to the publicly available  data,  various socio-economic parameters of all the districts of India have been selected for our study. 
In particular, for a wide spread analysis, we have chosen 14 socio-economic indicators (SEIs) from sectors such as education, employment, housing, health, etc.,
\textcolor{black}{keeping in mind the diversity of the indicators in different domains of life as well the availability of reliable data}.  
We refer to these indicators as Total Literate, Informal-Literate, High-Literate, Not-Literate, 
Main-Worker, Marginal-Worker, Workplaces (offices, factory, workshop, work-shed, etc.), Residence-Places, Market-Places, Education-Places (school, colleges, etc.), 
Tourist-Places, Medical-Places (hospital, etc.), High-Literate-Women, Main-Worker-Women \cite{censusdef}. 
Explicit definitions of all these SEIs are  provided  in Appendix \ref{APP:appndx}.

\subsection{Classification of Indian districts according to their urbanization status}
\label{SEC:classification}
\noindent 
 Keeping in mind the wide inter-state disparity in the urbanization level across the country, in this  new prospect  of scaling analysis at the district level, corresponding  to both the rural and urban population,  we would like to appropriately classify the districts in  homogeneous classes with respect to their urbanisation status. 
To start with, we first characterize the urbanisation in any district of India by the ratio of the urban and rural population in that district,  which we call the Urban-Rural ratio (URR). Based on this urbanization index URR, all the districts of India may be divided  into 4 distinct classes. Naturally,  the small values of URR will tune with  the Rural Class and the districts with URR $ >1$  will correspond  to the Urban Class.  The interim values of URR may be termed. as Semi- Rural ans Semi-Urban districts. 
\noindent 
For our analysis, we have distinctly classified  the rural, semi-rural, semi-urban and urban districts as:   
\begin{enumerate}
	\item Class-A (Rural): Rural districts are represented by URR  in between 0 and 0.1.  
	In this class, there are 108 and 126 districts in the year 2011 and 2001, respectively. 
	
	\item Class-B (Semi-rural): URR of these districts are in between 0.1 and 0.3. There are 271 districts in this class in 2011 and 242 districts  in 2001. 
	
	\item Class-C (Semi-urban): The semi-urban districts correspond to URR  in between 0.3 and 1.  
	The census data show that  there are  180  and 167 semi-urban districts in the years 2011 and 2001, respectively. 
	
	\item Class-D (Urban): URR of these districts are above 1 and are classified as urban districts of India. 
	Accordingly,  there are 81  and 55 urban districts in 2011 and 2001 respectively. 
\end{enumerate}

\begin{figure}[h]
\includegraphics[width=1.08\textwidth]{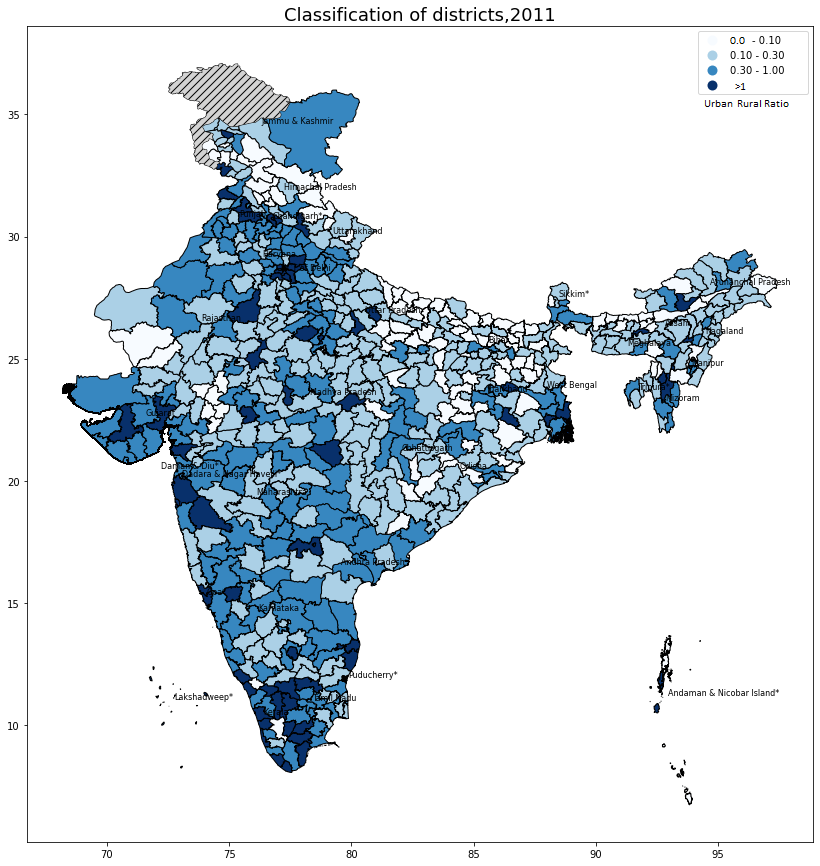}
\caption{Spatial distribution of 4 distinct classes (classified according to the urbanization index URR )  of Indian districts in the year 2011-Rural (108), Semi-Rural (271), Semi-Urban (180) and Urban (81).} %,  districts are  .} 
%  White- Rural districts (108), Sky-blue-Semii-Rural (271), Blue- Semi-Urban (180) and Navy-Blue- Urban (81)} 
\label{FIG:2011india}
\end{figure}

\noindent
See Figure \ref{FIG:2011india} for a spatial distributions of these four urbanisation classes in the year 2011. 
\textcolor{black}{We may point out here  that the cut-offs 0.1 and 0.3 used in the classification of districts are rather ad hoc;  but they lead to the desired homogeneity within each of the resulting four classes 
in respect to the urban scaling as we will see in our subsequent analyses. }

\noindent 
\begin{remark}: It has  been verified that the main conclusions about the nature and validity of the scaling law across districts within 
these four urbanisation classes do not differ significantly if these cut-offs are changed a little. 
\end{remark}

\newpage 

\newpage 
%\subsection{Fitting the Scaling Law to the Empirical Data}
\subsection{From the Empirical Data  to the  Scaling Law}
\label{SEC:Modelfit}

\noindent
The urban scaling law encompasses  a wide range of complex urban phenomena, including city fractals, Zipf's distribution, allometric growth, 1/f noise, power-law distance decay, scale-free network, and self-similar hierarchy of cities with cascade structure. [See for example \cite{chen_cata}].  %However, Dr. L.M.A. Bettencourt replaced all types of urban scaling laws with the allometric scaling law of cities, which led many young scholars to equate allometric scaling law with urban scaling law. 
However, for our analysis we use the common model,  the allometric urban scaling law \cite{chen_allo, lee, betten_2019_geo} that  is mathematically represented as 
\begin{equation}
Y_i=Y_0N_{i}^{\beta}e^{{\zeta}_i}, 
\label{EQ:ScalingLaw}
\end{equation}
where $N_{i}$ represents the population of the $i$-th district of a given urbanisation class at a given time point (2001 or 2011) 
and $Y_i$ represents the measure of a targeted SEI corresponding to that district.  
It follows from Eqn.~(\ref{EQ:ScalingLaw})  that there is a  linear relationship between  $\log Y_i$  and $\log N_i$.  
Thus, the scaling factor $\beta$ measures the average relative change of $Y_i$ with respect to $N_i$ in the logarithmic scale,
%, that is  $\langle\frac{ln(Y_i)}{ln(N_i)}\rangle $  % After taking log in both side of , it follows  a linear relationship between  $\log Y_i(t)$ and $\log N_i(t)$.{ln(N_i)}>$ 
and ${\zeta}_i$ measures the \textcolor{black}{deviance} of the $i$-th individual district from 
the scaling law (average relationship pattern) of the corresponding class.   
\\
\noindent
Based on our empirical data on population and any given SEI, 
we fit the scaling law (\ref{EQ:ScalingLaw}) separately for the four urbanisation classes  and for  the years 2001 and 2011. 
For any given year and class, the corresponding values of the scaling factor $\beta$ is estimated by the ordinary least square (OLS) method
applied to the linear regression model of $\log Y_i$s  on $\log N_i$s with $i$ varying over all districts within that particular class. 
More precisely, the OLS estimate of $\beta$ is obtained as 
\begin{equation}
\widehat{\beta}=\frac{\langle\log Y_i \log N_i\rangle  - \langle\log Y_i\rangle \langle\log N_i\rangle}{\langle\log N_i^2\rangle - \langle\log N_i\rangle^2},
\end{equation}
where $\langle \cdot \rangle$ represents the expectation (mean) operator. 
The process also provides an estimate of $Y_0$ as
\begin{equation}
\log\widehat{Y}_0 =  \langle\log Y_i\rangle - \widehat{\beta}\langle\log N_i\rangle
\end{equation}
and that of the individual \textcolor{black}{deviance} $\zeta_i$ as 
\begin{equation}
\widehat{\zeta}_i = \log Y_i - \log\widehat{Y}_0 - \widehat{\beta} \log N_i
\end{equation}
 for any $i$. 
The overall model error variance ($\sigma^2$) can then be estimated by the variance of the deviances
%, with standardization by appropriate degrees of freedom, 
as \begin{equation}
\widehat{\sigma}^2 = \frac{n}{n-2} \langle\widehat{\zeta}_i^2\rangle,
\end{equation}
where $n$ is the number of districts in that particular class.\\

\noindent 
{\bf Goodness-of-fit Measure}\\
After fitting the linear regression model, we  need  to determine how well our  model fits the data.  We use the  $R^2$  goodness-of-fit measure for linear regression models and test the appropriateness of the scaling law for the given empirical data.  The popular coefficient of variation (referred to as $R^2$) measure with higher values (close to one) indicates  better fit.
%Finally, the goodness for the fitted linear regression model, and hence the appropriateness of the scaling law for the given empirical data, 
%can be examined via the popular coefficient of variation (referred to as $R^2$) measure with higher values (close to one) indicating better fit.
The full estimation process is implemented by using the built-in functions in the software \texttt{MATLAB}. 

% under consideration.
%A particular deviance value beyond the range $[-1.96\widehat{\sigma}, 1.96\widehat{\sigma}]$ indicates that the corresponding district 
%has a significantly (at 95\% level under normality assumption) outlying behavior compared to the average scaling pattern of its class. 
%\noindent

%After fitting a linear regression model, one  needs to determine how well the model fits the data;  $R^2$  is a goodness-of-fit measure for linear regression models.
%Finally, the goodness for the fitted linear regression model, and hence the appropriateness of the scaling law for the given empirical data, 
%can be examined via the popular coefficient of variation (referred to as $R^2$) measure with higher values (close to one) indicating better fit.
%The full estimation process is implemented by using the built-in functions in the software \texttt{MATLAB}. 

{\color{black}
\begin{remark}
We would like to mention that, Shalizi \cite{Shalizi:2011} proposed using the per capita measures instead of the raw (extensive) values for the SEIs, 
which automatically show a ``mass" effect. If we use the per capita values of the SEIs ($Y_i/N_i$) in place of $Y_i$ in Equation (\ref{EQ:ScalingLaw}),
the resulting form of the scaling law would be 
\begin{equation}
\frac{Y_i}{N_i}=Y_0N_{i}^{\gamma}e^{{\zeta}_i}.
\end{equation}
By comparing it with Equation (\ref{EQ:ScalingLaw}), it can easily be seen that the scaling law in the per capita modeling must satisfy $\gamma=\beta-1$. 
By the location equivariance of our estimation method (OLS), the estimate of $\gamma$ would then be  $\widehat{\beta}-1$
and all subsequent inference would be equivalent. Only the linearity of the scaling exponent ($\beta=1$) in our formulation (\ref{EQ:ScalingLaw}) 
would be equivalent to the zero value (non-significance) of the scaling exponent in the per-capita formulation of the problem. 
Due to such equivalence, we have decided to continue with the well-known form of the scaling law as given in (\ref{EQ:ScalingLaw}) in our present paper. 
\end{remark}
}

\subsection{Scaling Exponent and Statistical Hypothesis Testing }%: Significance Testing on the Scaling Exponent}
\label{SEC:test}

%\noindent
%Hypothesis testing is a systematic method for deciding whether a parameter (e.g., scaling factor) in a population supports 
%a specific theory. %  (e.g., linear/sub-linear/super-linear) based on our empirical (sample) data. 

%\bigskip
\noindent 
Hypothesis testing is a systematic method for deciding whether a parameter (e.g., scaling factor) in a population supports 
a specific theory %  (e.g., linear/sub-linear/super-linear) 
based on our empirical (sample) data.\\
\noindent 
When there is an ambiguity in taking  a decision, statistical tests give a quantitative indication helping to deduce a conclusion with certain level (often 95\%) of confidence.  In the present context,  two types of hypothesis will be considered to (i)  test the linearity of the scaling factor and to (ii) test  the equality of the scaling factors .%we consider two types of hypothesis related to the scaling factors. 
\\

\noindent\textbf{Test for linearity of Scaling factor: }\\
Firstly, we consider  the null hypothesis that a given SEI (say $Y_i$) scale linearly, i.e., $H_0:{\beta}=1$ for the scaling factor $\beta$ of that particular class,
which we statistically test against  the omnibus alternative $H_1: {\beta}_1 \neq 1$.
For this purpose, we use the test statistics ($t_{\rm stat}$) computed as 
\begin{equation}
	t_{\rm stat}=\frac{\widehat{\beta} - 1}{SE(\widehat{\beta})},
\end{equation}
where the standard error of the estimated scaling factor $\widehat{\beta}$ is given by 
\begin{equation}SE(\widehat{\beta})=\widehat{\sigma}\left[\langle\log N_i^2\rangle - \langle\log N_i\rangle^2\right]^{-1/2}
\end{equation}
The probability of getting the t-value equal or greater than $t_{\rm stat}$ is known as %by  
the  p-value, which can be easily computed under normality assumption (that holds for classes with large number of districts). 
 The greater the value of $t_{\rm stat}$, less the p-value is and, hence, the more affirmation against null hypothesis is obtained. 
More precisely, if the resulting p-value is less than 0.05, we reject the null hypothesis (linearity of the scaling factor) at 95\% level of significance and infer the scaling law  to  follow a positive or negative scaling relation according to the sign of $(\widehat{\beta}-1)$. The scaling is termed as super-linear for $\beta >1$ and sub-linear for $\beta < 1$.% according to the sign of $(\widehat{\beta}-1)$. 

\medskip
\noindent\textbf{Test for equality of Two Scaling factors: }\\
The other interesting aspect is to compare the scaling law fitted to two classes (or at two time points) 
by testing for the equality of their respective scaling factors, say $\beta_1$ and $\beta_2$.  
The corresponding null and alternative hypotheses would be $H_0:{\beta}_1={\beta}_2$ and $H_1:{\beta}_1\neq {\beta}_2$, respectively.
Let us denote the estimated  scaling factors in the two classes (or two time-points) by $\widehat{\beta}_1$ and $\widehat{\beta}_2$, respectively, 
which are obtained based on the data on $n_1$ and $n_2$ districts in the respective classes. 
For this testing procedure, we use the two-sample test statistics given by 
\begin{eqnarray}
	t_{\rm stat}^{(2)} = \frac{\widehat{\beta}_2 -\widehat{\beta}_1}{\sqrt{ SE(\widehat{\beta}_1)^2+SE(\widehat{\beta}_2)^2}}, 
\end{eqnarray}
where the standard errors (SEs) are obtained as before from the empirical data of the respective classes. 
Under appropriate distributional assumptions, the statistics $t_{\rm stat}^{(2)}$ follows a $t$-distribution with degrees of freedom 
$df= \frac{ (SE(\widehat{\beta}_1)^2+SE(\widehat{\beta}_2)^2)^2}{\frac{SE(\widehat{\beta}_1)^4}{n_1}+\frac{SE(\widehat{\beta}_2)^4}{n_2}}$,
which helps us to compute the corresponding p-values. We infer that the two classes behave according to significantly different scaling rules (at 95\% level)
if the resulting p-value is obtained to be less than 0.05. 

\noindent For both the above-mentioned testing procedures, computation of the p-values is done using the software \texttt{MATLAB}.

% \begin{tabbing}
% \centering
% \begin{tabular}{r|rrr}
%       & Class-1 &   &Class-2 \\
%       \hline
% Slope($\beta$)  & ${\beta}_1$ &   & $ {\beta}_2$ \\
% No. of data points(n) & $n_1$ &   &$n_2$ \\
%  &   &    & \\
% Standard Error of regression $S_{\beta}$ &$\sqrt{\frac{\sum(y_i-\hat{y})^2}{n_1-2}}$ &      &$\sqrt{\frac{\sum(y_i-\hat{y})^2}{n_2-2}}$ \\
%    &   &    & \\
% Standard error of slope  SE(${\beta})$ &$\frac{S_{\beta 1}}{\sqrt{\sum(x_i-<x>)^2}} $&     &$\frac{S_{\beta 2}}{\sqrt{\sum(x_i-<x>)^2)}} $ \\
%  &   &    & \\
%Difference in slope(d) & \multicolumn{2}{c}{$ |{\beta}_2 -{\beta}_1|$} & \\
%\end{tabular}
%\label{tab:pval}	
% \end{tabbing}
%P value is calculated using Matlab function 

\section{Results and Analysis}
\label{SEC:reslt}

\subsection{Allometric Scaling Laws for the Rural and Urban  Districts }%On the existence of scaling law at district levels}

\noindent 
Following the methods described in Sec.II we have analysed the scaling features associated with various SEIs for both the years 2011 and 2001. Firstly,  we have estimated the values of the scaling exponents for the pool of all Indian districts (640 in the year 2011 and 590 in the year 2001) and then  for the four  distinct classes- rural, semi-rural, semi-urban and urban separately.  The estimated values of the (district level) scaling factors for each of the four urbanisation classes are presented in Table \ref{TAB:Scaling_Factor},
along with the same obtained for the pool of $all $ $districts$ of India, for both the years 2011 and 2001. 
Further, as a measure of the goodness-of-fit for the respective scaling laws %to the empirical data 
of  all the SEIs in different classes,  the corresponding $R^2$ values are also presented in the same Table \ref{TAB:Scaling_Factor}. 
See Figure \ref{FIG:Scaling2} below and Figures S1-S3 in the Supplementary Material for graphical representations of the fitted Scaling Laws 
for different SEIs in the years 2001 and 2011.

%==============================Table=============================
\begin{table}[!h]
	\caption{Estimated scaling factors ($\beta$) for all SEIs at the district level, within four urbanisation classes (A--D) and also across all India.
	The $R^2$ values indicating the goodness-of-fit for the corresponding scaling law is given in the respective parenthesis.}
	\centering	
%	\resizebox{0.8\textwidth}{!}{
		\begin{tabular}{l|l|llll}\hline
	&	All India	&	Class A	&	Class B	&	Class C	&	Class D	\\
SEI	&	All Districts	&	Rural	&	Semi-rural	&	Semi-urban	&	Urban	\\\hline\hline
			&	\multicolumn{5}{|c}{\textbf{Year 2011}}\\
Total-Literate	&	1.001 (0.97)~~	&	0.955 (0.97)~~	&	0.995 (0.98)~~	&	0.981 (0.99)~~	&	0.987 (1.00)~~	\\
Informal-Literate 	&	1.103 (0.87)	&	1.136 (0.90)	&	1.068 (0.83)	&	1.123 (0.86)	&	1.068 (0.91)	\\
High-Literate	&	1.117 (0.82)	&	0.929 (0.82)	&	1.076 (0.88)	&	1.051 (0.92)	&	1.088 (0.92)	\\
Not-Literate	&	1.005 (0.93)	&	1.072 (0.95)	&	1.012 (0.95)	&	1.045 (0.95)	&	1.042 (0.96)	\\
Main-Workers	&	0.964 (0.94)	&	0.856 (0.94)	&	0.934 (0.93)	&	0.982 (0.96)	&	1.018 (0.98)	\\
Marginal-Worker	&	0.933 (0.77)	&	1.015 (0.90)	&	0.989 (0.86)	&	0.994 (0.86)	&	1.004 (0.84)	\\
Workplaces		&	0.979 (0.70)	&	0.746 (0.66)	&	0.907 (0.74)	&	0.971 (0.77)	&	1.041 (0.77)	\\
Residence-Places	&	0.998 (0.98)	&	0.945 (0.97)	&	0.998 (0.97)	&	1.001 (0.98)	&	1.010 (0.98)	\\
Market-Places	&	0.987 (0.79)	&	0.795 (0.70)	&	0.978 (0.81)	&	0.973 (0.84)	&	0.953 (0.87)	\\
Education-Places	&	0.722 (0.76)	&	0.630 (0.75)	&	0.749 (0.78)	&	0.777 (0.81)	&	0.815 (0.82)	\\
Tourist-Places	&	0.751 (0.59)	&	0.510 (0.48)	&	0.701 (0.57)	&	0.828 (0.67)	&	0.714 (0.63)	\\
Medical-Places	&	0.955 (0.85)	&	0.704 (0.71)	&	0.947 (0.87)	&	0.976 (0.91)	&	1.046 (0.92)	\\
High-Literate-Women	&	1.119 (0.73)	&	0.858 (0.69)	&	1.047 (0.81)	&	1.052 (0.87)	&	1.095 (0.90)	\\
Main-Worker-Women	&	0.866 (0.71)	&	0.728 (0.69)	&	0.799 (0.63)	&	0.925 (0.73)	&	1.004 (0.87)	\\\hline
			&	\multicolumn{5}{|c}{\textbf{Year 2001}}\\
Total-Literate	&	0.993 (0.95)	&	0.903 (0.94)	&	0.999 (0.94)	&	0.982 (0.97)	&	0.989 (0.99)	\\
Informal-Literate 	&	1.084 (0.76)	&	1.066 (0.83)	&	1.077 (0.69)	&	1.168 (0.77)	&	1.076 (0.72)	\\
High-Literate	&	1.116 (0.81)	&	1.010 (0.87)	&	1.049 (0.86)	&	1.068 (0.89)	&	1.065 (0.87)	\\
Not-Literate	&	1.009 (0.93)	&	1.086 (0.97)	&	1.001 (0.92)	&	1.033 (0.95)	&	1.038 (0.97)	\\
Main-Workers	&	0.917 (0.90)	&	0.779 (0.85)	&	0.929 (0.92)	&	0.965 (0.96)	&	1.021 (0.84)	\\
Marginal-Worker	&	0.918 (0.77)	&	0.959 (0.88)	&	0.941 (0.82)	&	1.006 (0.88)	&	0.998 (0.77)	\\
Workplaces	&	1.003 (0.73)	&	0.818 (0.71)	&	0.973 (0.73)	&	0.988 (0.75)	&	1.064 (0.82)	\\
Residence-Places	&	0.986 (0.97)	&	0.950 (0.97)	&	0.995 (0.96)	&	0.991 (0.98)	&	1.000 (0.99)	\\
Market-Places	&	1.026 (0.79)	&	0.881 (0.76)	&	1.002 (0.78)	&	1.030 (0.85)	&	0.980 (0.89)	\\
Education-Places	&	0.738 (0.77)	&	0.629 (0.74)	&	0.779 (0.74)	&	0.819 (0.84)	&	0.794 (0.84)	\\
Tourist-Places	&	0.738 (0.55)	&	0.453 (0.41)	&	0.705 (0.45)	&	0.834 (0.67)	&	0.818 (0.75)	\\
Medical-Places	&	0.989 (0.86)	&	0.819 (0.77)	&	1.002 (0.84)	&	1.014 (0.92)	&	1.054 (0.96)	\\
High-Literate-Women	&	1.126 (0.68)	&	0.905 (0.73)	&	1.013 (0.73)	&	1.092 (0.81)	&	1.076 (0.85)	\\
Main-Worker-Women	&	0.781 (0.62)	&	0.604 (0.60)	&	0.747 (0.49)	&	0.866 (0.68)	&	0.945 (0.76)	\\
			\hline
	\end{tabular}
%}
	\label{TAB:Scaling_Factor}
\end{table}

\begin{figure}[!h]
	\centering
	\subfloat[Total-Literate]{
		\includegraphics[width=0.56\textwidth]{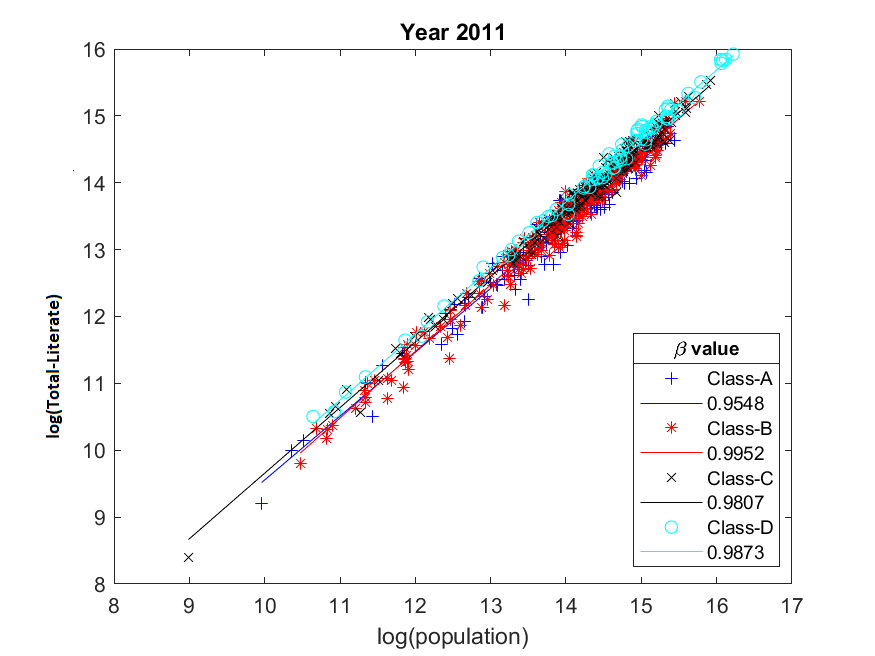}
		\includegraphics[width=0.56\textwidth]{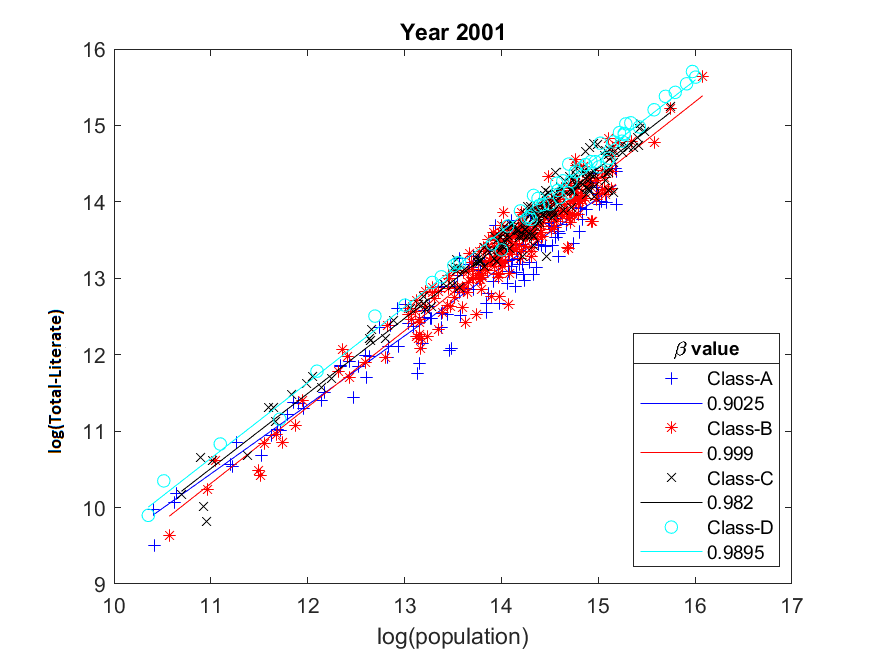}
		\label{FIG:Total-Literate}}
	%~ %--------------------------------------------------------------------
	%	\subfloat[\large $2001$]{
	%		\includegraphics[width=0.50\textwidth]{scaling_2001_y1.png}
	%		\label{FIG:2001var1}}
	\\
	\subfloat[Main-Workers]{
		\includegraphics[width=0.56\textwidth]{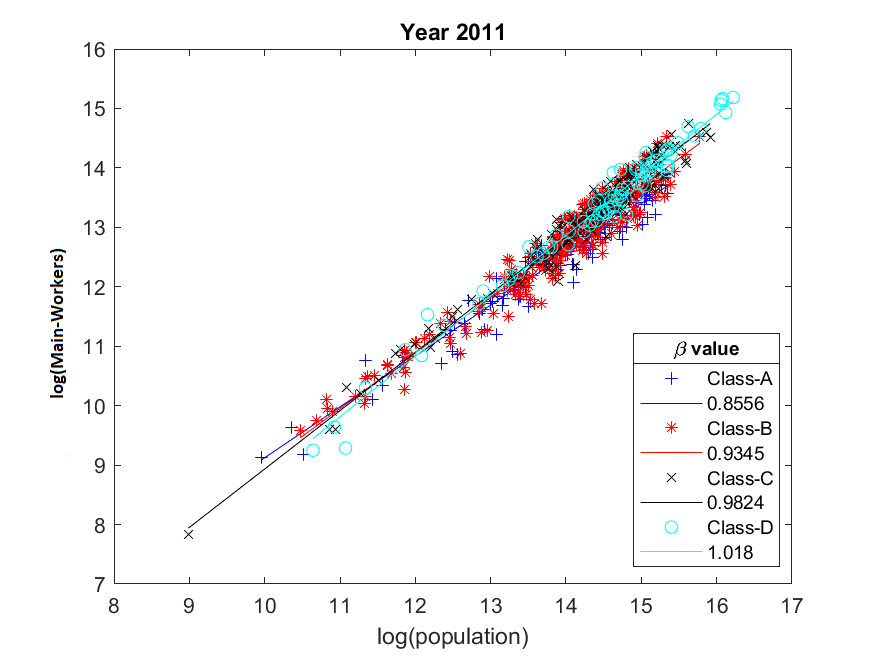}
		\includegraphics[width=0.56\textwidth]{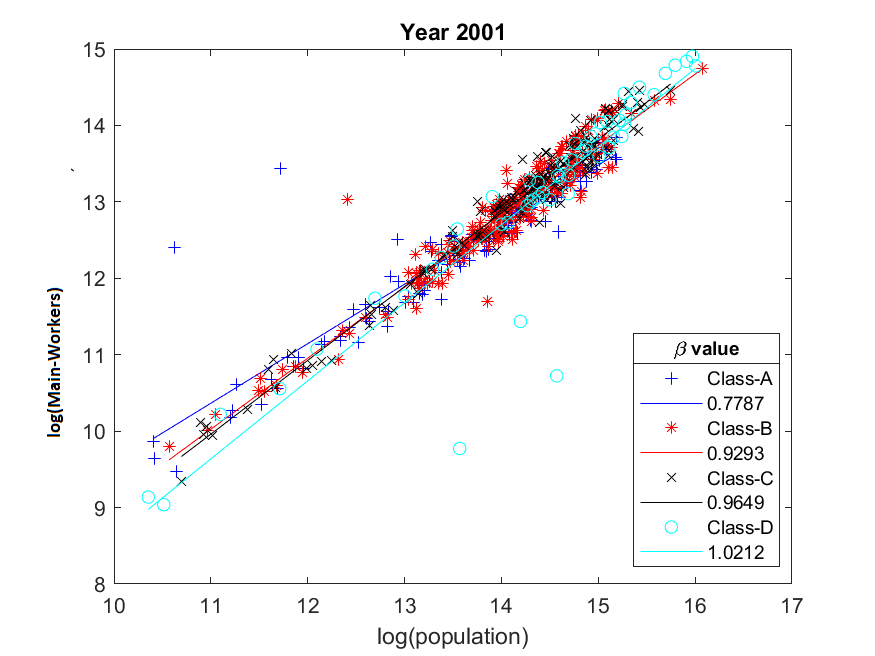}
		\label{FIG:Main-Workers}}
	%~ %--------------------------------------------------------------------
	%	\subfloat[\large $2001$]{
	%		\includegraphics[width=0.50\textwidth]{scaling_2001_y2.png}
	%		\label{FIG:2001var1}}
	\caption{Scaling laws fitted to the district level data within the four urbanisation classes (A--D), for both the years 2011 and 2001. (a) Plots for Total-Literate and (b) Plots for Main-Workers . Respective $\beta$ values are  provided in the insets. } %fitted to the district level data within the four urbanisation classes (A--D), for both the years 2011 and 2001.}
	\label{FIG:Scaling2}
\end{figure}

\noindent
It is remarkable  to note that our result shows the existence of  the allometric scaling laws  even at the {\it{ level of Indian districts}}  for all education related SEIs 
(Total-Literate, Informal-Literate, High-Literate, Not-Literate) as well as for the Main-Workers, Residence-Places and Medical-Places; 
the $R^2$ values obtained for these SEIs are very high (often greater than 0.8 or 0.9) in both the years 2011 and 2001.  
More interestingly, the scaling law behaves in a robust manner for all these SEIs (except for Medical-Places) yielding high $R^2$ 
(and hence greater validity for the fitted scaling law) at all the four urbanisation classes. 
However, class-wise scaling factors deviate from the average (all India level) scaling factor
with their values being below the average in rural areas (Class A) for most SEIs.

\begin{figure}[h]
	\includegraphics[width=0.80\textwidth]{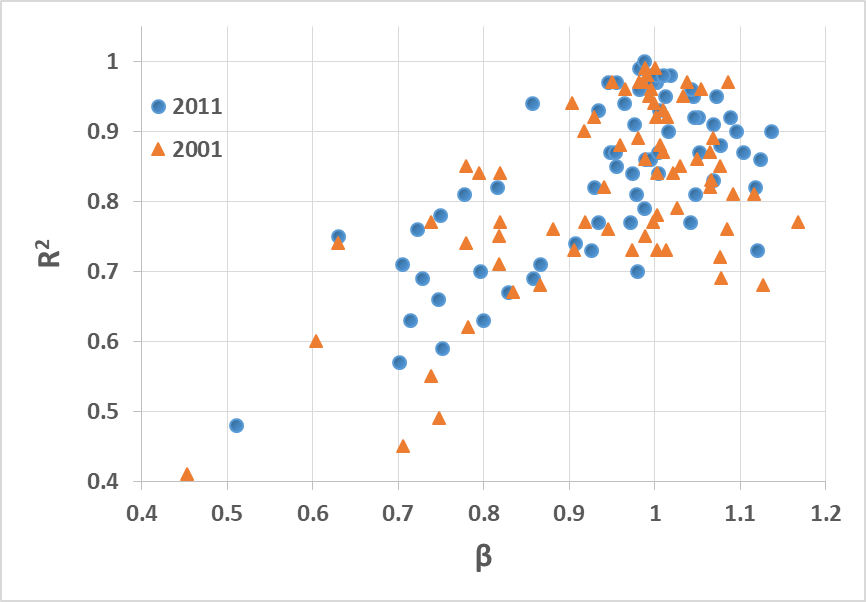}
	\caption{Plot of the goodness-of-fit measures ($R^2$) vs.~the estimated scaling exponent ($\beta$) for the years 2001 and 2011.  $R^2$ values are higher ( depicting  good fit ) for values of $\beta$ close to 1} %. : comparative representation  for the years 2001 and 2011. }
	\label{FIG:R2_beta}
\end{figure}

\noindent
For the  SEIs, for which the validity of the scaling law is not very satisfactory (slightly lower $R^2$ values) at all India districts level, 
the existence of the scaling law becomes more prominent with greater validity (higher $R^2$ values) 
as we restrict ourselves within the urbanisation classes with greater urban populations. 
In particular, among the urban districts (Class D), 
all our SEIs (except only for Tourist-Places) exhibit strong evidences (higher $R^2$ values greater than 0.8) of the existence of scaling law in 2011.
The values of $R^2$ was a bit lower for these SEIs in the year 2001. Thus, in general, the existence of scaling law  in all Indian district level data 
is seen to become more prominent over time, and also for classes with higher URR at any given time-point. 
Particularly at the year 2011, the validity of scaling law weakens considerably as we move continuously from Class D to Class A (urban to rural) 
for High-Literate-Women, Main-Worker-Women and all SEIs related to places; their $R^2$ values indeed become less than 0.8 in Class A, 
except for the Residence-Places that still has high  $R^2=0.97$ in Class A.   
An interestingly opposite trend is observed with Main-Workers, for which the degree of validity towards scaling law 
decreases from Class A (with $R^2=0.9$) to Class D ($R^2=0.84$).
\textcolor{black}{Figure \ref{FIG:R2_beta} shows the plot of the $R^2$-values as a function of the estimated scaling exponent for both the years 2001 and 2011;
we can clearly see that, in most cases, the fits of the scaling law are better, having a higher $R^2$ on average than others, 
for the SEIs which yield scaling exponents close to 1.
}

\noindent
We may note the change in the $\beta$  values corresponding to  the various SEIs in a decade (2001 to 2011) from the Table  \ref{TAB:Scaling_Factor}.  
Interestingly, within this period, for  most of the variables the change in  $\beta$ is very small. 
To be specific, the changes in $\beta$ are in the range [0.002 - 0.059] only within the urban class (Class D) except for the Tourist Places  where the change is 0.104. 
For the Class C districts,   the change in $\beta$ is within the range of [0.001 - 0.059], whereas  it is within [0.004 - 0.055] for the Class B.  
For the Rural population (Class A) the change in $\beta$ is a little heterogeneous;  the range is [0.052 - 0.124]  for most of the variables 
except for the Not literate, Residence Places, Education Places and Tourist Places where the change is small, within the range [0.001 - 0.014] only.

\subsection{Linearity of the scaling index: a statistical test}
\noindent
It is stimulating to  statistically examine the values of the estimated scaling exponents  from Table \ref{TAB:Scaling_Factor} to see 
if the corresponding SEI deviates from the linear scaling law ($\beta=1$) and follows a sub-linear (($\beta < 1$). or super-linear ($\beta  > 1$) pattern. 
%(according to whether the scaling factors are less or greater than one, respectively). 
Actually, the $super$-l$inear$ allometric relation  corresponds to positive allometric scaling relation, whereas  the  $sub$-$linear$  allometric relation corresponds to the  negative allometric scaling relation.  The linearity  of the scaling exponent is. sometimes  termed as $isometric$  $growth$  in the  literature.  (See for example, \cite{lee, chen_2}).

\noindent 
To adjust for the sampling fluctuations near the value one, we have statistically tested if the estimated scaling factor deviates significantly from one (at 95\% level).
Consistently, we comment on the nature (or behavior) of the scaling law for all  SEIs in different urbanisation classes
which are reported in Table \ref{TAB:Scaling_Compare_linear};
the exact p-value for each test are provided in Table S1 of the Supplementary material. 
\textcolor{black}{Recall that these p-values are obtained from a statistical test for the null hypothesis of linearity of the scaling exponent 
	(as described in Section \ref{SEC:test}), and hence, smaller the p-values we have larger evidence against the null hypothesis (linearity) 
	so that we reject it in favour of non-linearity of the scaling exponent.
}

%==============================Table=============================
\begin{table}[!h]
	\caption{Results from the statistical tests for linearity of the  scaling factors at 95\% level of significance.}
	\centering	
\resizebox{\textwidth}{!}{
	\begin{tabular}{l|llll|llll}\hline
SEI	&	Class A	&	Class B	&	Class C	&	Class D	&	Class A	&	Class B	&	Class C	&	Class D	\\
		&	\multicolumn{4}{|c}{\textbf{Year 2011}}	&	\multicolumn{4}{|c}{\textbf{Year 2001}}\\\hline
Total-Literate	&	Sub-linear	&	\textbf{Linear}	&	Sub-linear	&	\textbf{Linear}	&	Sub-linear	&	\textbf{Linear}	&	\textbf{Linear}	&	\textbf{Linear}	\\
Informal-Literate	&	\it{Super-linear}	&	\it{Super-linear}	&	\it{Super-linear}	&	\textbf{Linear}	&	\textbf{Linear}	&	\textbf{Linear}	&	\it{Super-linear}	&	\textbf{Linear}	\\
High-Literate	&	\textbf{Linear}	&	\it{Super-linear}	&	\it{Super-linear}	&	\it{Super-linear}	&	\textbf{Linear}	&	\textbf{Linear}	&	\it{Super-linear}	&	\textbf{Linear}	\\
Not-Literate	&	\it{Super-linear}	&	\textbf{Linear}	&	\it{Super-linear}	&	\textbf{Linear}	&	\it{Super-linear}	&	\textbf{Linear}	&	\textbf{Linear}	&	\textbf{Linear}	\\
Main-Workers	&	Sub-linear	&	Sub-linear	&	\textbf{Linear}	&	\textbf{Linear}	&	Sub-linear	&	Sub-linear	&	Sub-linear	&	\textbf{Linear}	\\
Marginal-Worker	&	\textbf{Linear}	&	\textbf{Linear}	&	\textbf{Linear}	&	\textbf{Linear}	&	\textbf{Linear}	&	Sub-linear	&	\textbf{Linear}	&	\textbf{Linear}	\\
Workplaces	&	Sub-linear	&	Sub-linear	&	\textbf{Linear}	&	\textbf{Linear}	&	Sub-linear	&	\textbf{Linear}	&	\textbf{Linear}	&	\textbf{Linear}	\\
Residence-Places	&	Sub-linear	&	\textbf{Linear}	&	\textbf{Linear}	&	\textbf{Linear}	&	Sub-linear	&	\textbf{Linear}	&	\textbf{Linear}	&	\textbf{Linear}	\\
Market-Places	&	Sub-linear	&	\textbf{Linear}	&	\textbf{Linear}	&	\textbf{Linear}	&	Sub-linear	&	\textbf{Linear}	&	\textbf{Linear}	&	\textbf{Linear}	\\
Education-Places	&	Sub-linear	&	Sub-linear	&	Sub-linear	&	Sub-linear	&	Sub-linear	&	Sub-linear	&	Sub-linear	&	Sub-linear	\\
Tourist-Places	&	Sub-linear	&	Sub-linear	&	Sub-linear	&	Sub-linear	&	Sub-linear	&	Sub-linear	&	Sub-linear	&	Sub-linear	\\
Medical-Places	&	Sub-linear	&	Sub-linear	&	\textbf{Linear}	&	\textbf{Linear}	&	Sub-linear	&	\textbf{Linear}	&	\textbf{Linear}	&	\textbf{Linear}	\\
High-Literate-Women	&	Sub-linear	&	\textbf{Linear}	&	\textbf{Linear}	&	Sub-linear	&	\textbf{Linear}	&	\textbf{Linear}	&	\it{Super-linear}	&	\textbf{Linear}	\\
Main-Worker-Women	&	Sub-linear	&	Sub-linear	&	\textbf{Linear}	&	\textbf{Linear}	&	Sub-linear	&	Sub-linear	&	Sub-linear	&	\textbf{Linear}	\\
		\hline
	\end{tabular}
	}
	\label{TAB:Scaling_Compare_linear}
\end{table}

\noindent
It is to be noted from Table \ref{TAB:Scaling_Compare_linear} that  %the scaling law suggests that 
in the urban Class D, most of the SEIs corresponding to individual basic needs (education, house) behave $linearly$  with respect to the population  
in both the years 2011 and 2001, except for Education-Places, Tourist-Places, High-Literate and High-Literate-Women. However, Education-Places and  Tourist-Places are seen to follow sub-linear scaling relationship in both the years 2001 and 2011.
Interestingly, the scaling factor for %the last  two  SEIs 
SEI  related to higher education, %were indeed 
linear in the year 2001 but changes in 2011. %, where High-Literate indicator becomes super-linear and High-Literate-Women indicator features a  sub-linear pattern. It may be noted that the   estimated scaling factor, corresponding to High -Literate and High-Literate-Women, at all India level is observed to be around $\beta\sim 1.12$ for both the years 2001 and 2011, implying a  good sign for urban development. 
\noindent 
SEIs connected to economic activities in a country, namely  Main-Workers, Marginal-Workers and Workplaces scale linearly in urban Class D and  becomes sub-linear as the proportion of rural population increases in our urbanisation classes; 
Main-Workers and Workplaces indeed scales sub-linearly within the rural Class A in both the years. 
It is good to note that the sub-linear scaling relationship of the variable Main-Workers in 2001 has become linear in 2011. 
%Employment is a  big issue in developing countries. To battle the current economic crisis India should have proper plan about  employment which will enhance labour supply. 

%\newpage 
%\subsection{Results from Gender perspective}

\noindent 
 Tables  \ref{TAB:Scaling_Factor} and \ref{TAB:Scaling_Compare_linear} indicate  that the allometric  scaling analysis in all the four urbanization classes have been analysed  from the gender perspective also. %For this purpose, we have considered data for the High-Literate-Women and Main-Worker-Women to represent SEIs  from the education and economic sectors.
The last 2. rows of Table \ref{TAB:Scaling_Compare_linear}  depicts  the  nature  of the linearity of the scaling factors for the  % from the gender perspective. We may note  the changes in the behaviour, in terms of linearity,  of the scaling factors for %the last  two  SEIs 
SEI s related to higher education in Women  and Women Main workers  %were indeed 
corresponding to  the  years  from 2001 to 2011 in all the 4 urbanization classes.. %, where High-Literate indicator becomes super-linear and High-Literate-Women indicator features a  sub-linear pattern.
The Table \ref{TAB:Scaling_Factor} %and  \ref{TAB:Female_Scaling_Compare_linear} 
shows that the   estimated scaling factors, corresponding to High -Literate and High-Literate-Women, at all India level is observed to be around $\beta\sim 1.12$ for both the years 2001 and 2011, implying a  good sign for urban development. 

%\begin{table}[!h]
%	\caption{Results from the statistical tests for linearity of the  scaling factors at 95\% level of significance.}
%	\centering	
%\resizebox{\textwidth}{!}{
%	\begin{tabular}{l|llll|llll}\hline
%SEI	&	Class A	&	Class B	&	Class C	&	Class D	&	Class A	&	Class B	&	Class C	&	Class D	\\
%		&	\multicolumn{4}{|c}{\textbf{Year 2011}}	&	\multicolumn{4}{|c}{\textbf{Year 2001}}\\\hline
		
%		High-Literate-Women	&	Sub-linear	&	\textbf{Linear}	&	\textbf{Linear}	&	Sub-linear	&	\textbf{Linear}	&	\textbf{Linear}	&	\it{Super-linear}	&	\textbf{Linear}	\\
%Main-Worker-Women	&	Sub-linear	&	Sub-linear	&	\textbf{Linear}	&	\textbf{Linear}	&	Sub-linear	&	Sub-linear	&	Sub-linear	&	\textbf{Linear}	\\
%		\hline
%	\end{tabular}
%	}
%	\label{TAB:Female_Scaling_Compare_linear}
%\end{table}

\noindent It is understood that the super-linear scaling ( positive allometric scaling) represents faster  increment  of the indicators with respect to the  system size while sub-linear scaling ( negative allometric scaling) indicates  slower returns. The results in the Tables \ref{TAB:Scaling_Compare_linear} and %\ref{TAB:Female_Scaling_Compare_linear} 
may help  the policy makers to build  the  strategies for the betterment of  the rural, semi-rural, semi-urban and urban population of the Indian districts. Care may also be taken to improve the  infra-structural facilities in the rural India. For example, care may be taken so that  the super-linear scaling of Not Literate variable for the Class A people at least becomes linear. As the number of main-workers and number of factories/workshops are correlated with each other, policy makers should encourage  for more industry (Workplaces which is now sub-linear) in rural areas.

\subsection{Comparison of the scaling exponents across different classes}

\noindent
We compare the scaling factors of various SEIs of  the four  distinct  urbanisation classes using appropriate statistical tests 
and the findings obtained at 95\% level of significance 
(Detailed results  reported in Table \ref{TAB:Scaling_Compare_Classes}, Appendix  B). 
%\ref{TAB:Scaling_Compare_Classes}. 
The exact p-values of all the pairwise comparison tests are provided in Table S2 of the Supplementary Material.

%==============================Table=============================

\noindent We note that, most interestingly, there is no significant difference between the scaling factors obtained in Class C (semi-urban) and Class D (urban)  
for all SEIs indicating that they behave similarly in both semi-urban and urban classes; this is indeed true for both the years 2001 and 2011. 
While comparing with  rural and semi-rural classes as well, we see that all the four classes have statistically similar scaling law behaviours (no significant differences)
in terms of Informal-Literate and Marginal-Workers in both the years (2001 and 2011). 
There was a statistically significant difference of the rural Class A from the rest of the classes (B, C, D) in terms of Total-Literate in 2001
but that difference has almost vanished in the year 2011. On the contrary, 
there was no significant difference in the scaling law of High-Literate across the urbanisation classes in 2001, 
but a significant discrimination has been developed in 2011 departing the rural Class A from the rest. 
These indicate that, over the 10 years (from 2001 to 2011),  there has been  changes in the number of highly educated people in rural India, 
which needs to be investigated further for proper policy making. 
The slight differences in the scaling behavior of Not-Literate across the urbanisation classes, however, seem to further decrease over time. 

\noindent
The scaling exponents $\beta$ of the economical SEIs, Main-worker and Workplaces,  of the rural Class A is seen to differ significantly 
from the rest of the classes which are rather statistically similar in the year 2001. 
In the year 2011, however, the variable Main-Workers follows significantly different scaling law even between the rural and semi-rural classes 
but the same for Workplaces are still similar between these two classes . 
%It indicates that *********** any explanation?? *****
In terms of other infrastructural SEIs related to different types of places, there has been a clear discrimination between the scaling factors
of the rural class versus the rest (Classes B, C and D, which do not differ significantly among themselves) in both the years 2001 and 2011.

\section{Discussion and Conclusion  }
\label{SEC:conclusion}

\noindent

%The urban scaling law encompasses a wide range of complex urban phenomena, including city fractals, Zipf's distribution, allometric growth, 1/f noise, power-law distance decay, scale-free network, and self-similar hierarchy of cities with cascade structure. However, Dr. L.M.A. Bettencourt replaced all types of urban scaling laws with the allometric scaling law of cities, which led many young scholars to equate allometric scaling law with urban scaling law. \\
\noindent
In recent years, concept of scaling has become one of the attentive topics  in the area of urban-rural research. %In fact, scaling is not important, scaling law or scaling relation is important in urban or rural studies. 
The study of scaling law is   important as  it provides a new way to analyze scale-free systems. 
As a first attempt to formulate a unified framework  of $scaling$,  going beyond cities,  comprising both the rural and urban population,  we considered 
%understand the existence of scaling laws in 
larger district-like administrative units. 
This interdisciplinary work, a multi-scale analysis  based on  rural  and urban population, sheds light on the understanding of scaling laws  emerging  in Indian districts.  Here, we have investigated the existence of allometric scaling laws at the district (second tier administrative units) level  corresponding to the urbanization  in India.
We study the allometric scaling relations of several socio-economic indicators corresponding to the population of all Indian districts as well as  the population   (referred to size of the class) of  the four distinct classes- rural, semi-rural, semi-urban and urban  classes of Indian districts  separately for the years 2001 and 2011.
Adopting  a rigorous statistical hypothesis  test  we  check   the linearity of the corresponding scaling exponents $\beta$.   The linearity test helps us  to inspect whether the estimated scaling index ($\beta$) for any particular socio-economic indicator or infra-structural  variable behaves in a linear,  sub-linear or super-linear manner indicating the rates of growth of the underlying factor with the population of the districts within a given urbanisation class.

\noindent
{\textcolor{black}{  
%As a first attempt to formulate a unified framework  of $scaling$,  comprising both the rural and urban population,  we considered 
%understand the existence of scaling laws in 
%larger district-like administrative units. 
We have classified all Indian districts in four groups (URR classes) based on the proportion of urban to rural populations within the districts. We then  investigated the scaling laws of all the Indian districts as well as within such groups so that the analysis may help to  reduce the heterogeneity between districts in terms of their urbanization structure.  The scaling exponents, associated  with various SEIs of the rural, semi-rural, semi-urban and urban population  classes have been estimated. Subsequently,  the  linearity  of the scaling exponent  has been inspected  through a statistical hypothesis testing. Moreover, to understand the change in  the scaling behaviour with time, we have  presented  a comparative analysis with the data for the years 2001 and 2011.  Our results indicate that not all the scaling factors behave linearly, some of them characterize a super-linear ( a positive allometric relation) behaviour and some behave sub-linearly ( negative allometric relation). These results clearly illustrate that more investigation along this line would be really useful to further understand the underlying process of urbanizations within districts. As  India stands  today at the inflection point of urban transformation,  we believe, this data based comprehensive study of  allometric scaling in Indian districts will help in making the urban transformation more effective and accelerate  future work in this  area of rural-urban  research.}}

\noindent 
The present multi-scale analysis with rural and urban population may set-off future study  in the area of fractal analysis of urban scaling. %This work may trigger many future projects  in the area of rural-urban research. 
 With the model presented here, one can study the fractal/multi-fractal features  of  the allometric scaling exponents
\cite{chen_fractal, batty_fractal,chen_fractal2} associated with various SEIs of the rural and urban districts.
 Also  one can make quantitative predictions about the economic gain that can be expected with certain changes in various socio-economic variables and its population size.
This may be be aided by the  estimated super-linear and sub-linear scaling exponents appearing in various SEIs relative 
to urban, semi-urban, semi-rural and rural population of Indian districts and their changes over time.

\noindent Urban scaling laws are deeply related with the ways people  live in an area. Generally, %we find  
cities within a district, considered as areas of settlement of the urban population, have  many facilities of livelihood,  whereas semi-towns, towns or villages comprising  the increasing proportion of rural population  lack the requisite infrastructure. It is important to perceive  the  variation  in  different  rural- and urban- level  factors  with respect to the population-size to understand the underlying dynamics of the urbanization process..%  can be instrumental in  creation of a unified theory of scaling  in the rural and urban areas.
We believe that the present study reflects the actual  administrative/economic/physical activities of  rural and urban regions  and reveal the general characteristics of infrastructure and services in the urban as well as  rural areas of India.  This comparative scaling analysis of rural and urban India  within a unified framework, a  significant contribution in the literature of allometric urban scaling law, may,
%These results will, 
in turn, guide the policymakers to take appropriate measures to maintain sustainable developments of 
these important socio-economic factors  along with the continuous urbanisation of India.

\bigskip \noindent
{\bf{Acknowledgment}}:\\
%SC thanks Science and Engineering Research Board (SERB), Government of India for financial support through Junior Research Fellowship (Grant No.CRG/2019/001461). The 
This  research is supported by a research grant (No.~CRG/2019/001461) from the Science and Engineering Research Board (SERB), Government of India, India.  
The authors also thank the anonymous reviewers for their constructive suggestions.

%\newpage

\appendix
\section{Definitions of our Socio-Economic Indicators (SEIs)}
\label{APP:appndx}
\noindent 
{\small
In the following, we briefly describe the  definitions of all the socio-economic variables considered in our analysis \cite{censusdef}. 
\begin{itemize}
	\item Total Literate: Total number of literate persons. A person aged 7 years and above, who is able to both read and write, is defined as a literate person.  
	\item Informal-Literate: Number of literate persons who never go to any educational institute like school.
	\item High-Literate: Total number of persons, having a degree of graduation and above.  
	\item Not-literate: Total number of illiterates. 
	A person, who cannot read and write or can read but cannot write, is considered to be as illiterate along with all children of age on or bellow 6 years.  
	\item Main-Workers: Total number of main workers. A person working  for an economic productivity \cite{censusdef} 
	(production and self-consumption also considered as economic activity) for more than 6 months is defined as a main worker.  
	\item Marginal-Workers: Total number of marginal workers, defined as the person working for less than 6 months.  
	\item Workplaces: Total number of  Factory, workshop, workshed etc. 
	\item Residence-Places: Total number of census houses used only for residential purposes.  
	A `Census house' is a building or part of a building used or recognized as a separate unit 
	because it has a separate main entrance from the road or common courtyard or staircase etc. 
	It may be occupied or vacant and may be used for residential or nonresidential purposes or both.
	\item Market-Places:  Total number of census houses used only for business purposes like shops and offices.
	\item Education-Places:   Total number of school colleges, etc.
	\item Tourist-Places:  Total number of  Hotel, lodge, guest house, etc.
	\item Medical-Places:   Total number of  Hospital and Dispensary, etc
	\item High-Literate-Women: Total number of women, having a degree of graduation and above.   
	\item Main-Worker-Women: Total number of women who are main workers.  
%Total food production in whole year is defined as production....(?) .

\end{itemize}
}

%\vspace*{2cm}

%\newpage 

\begin{center}
{\bf{ Appendix B:  Comparison of estimated Scaling factors}}
\end{center}

\noindent
The following Table \ref{TAB:Scaling_Compare_Classes} shows the details of the comparison of the scaling factors for both the years.

\begin{table}[!h]
\caption{Results of the statistical tests to compare the estimated scaling factors between any two  urbanisation classes (among A, B, C, D).
	H1 indicates that the corresponding scaling factors are significantly different at 95\% level of significance; 
	whereas H0 indicates that there is no evidence to conclude their differences to be statistically significant.}
	
	\centering	
	%	\resizebox{0.8\textwidth}{!}{
	\begin{tabular}{l|rrrrrr|rrrrrr}\hline
SEI	&	A--D	&	A--C	&	A--B	&	B--C	&	B--D	&	C--D	&	A--D	&	A--C	&	A--B	&	B--C	&	B--D	&	C--D	\\\hline
&	\multicolumn{6}{|c}{\textbf{Year 2011}}	&	\multicolumn{6}{|c}{\textbf{Year 2001}}\\
Total-Literate	&	H0	&	H0	&	H1	&	H0	&	H0	&	H0	&	H1	&	H1	&	H1	&	H0	&	H0	&	H0	\\
Informal-Literate 	&	H0	&	H0	&	H0	&	H0	&	H0	&	H0	&	H0	&	H0	&	H0	&	H0	&	H0	&	H0	\\
High-Literate	&	H1	&	H1	&	H1	&	H0	&	H0	&	H0	&	H0	&	H0	&	H0	&	H0	&	H0	&	H0	\\
Not-Literate	&	H0	&	H0	&	H1	&	H0	&	H0	&	H0	&	H0	&	H1	&	H1	&	H0	&	H0	&	H0	\\
Main-Workers	&	H1	&	H1	&	H1	&	H1	&	H1	&	H0	&	H1	&	H1	&	H1	&	H0	&	H0	&	H0	\\
Marginal-Worker	&	H0	&	H0	&	H0	&	H0	&	H0	&	H0	&	H0	&	H0	&	H0	&	H0	&	H0	&	H0	\\
Workplaces	&	H1	&	H1	&	H1	&	H0	&	H0	&	H0	&	H1	&	H1	&	H1	&	H0	&	H0	&	H0	\\
Residence-Places	&	H1	&	H1	&	H1	&	H0	&	H0	&	H0	&	H1	&	H1	&	H1	&	H0	&	H0	&	H0	\\
Market-Places	&	H1	&	H1	&	H1	&	H0	&	H0	&	H0	&	H0	&	H1	&	H1	&	H0	&	H0	&	H0	\\
Education-Places	&	H1	&	H1	&	H1	&	H0	&	H0	&	H0	&	H1	&	H1	&	H1	&	H0	&	H0	&	H0	\\
Tourist-Places	&	H1	&	H1	&	H1	&	H1	&	H0	&	H0	&	H1	&	H1	&	H1	&	H0	&	H0	&	H0	\\
Medical-Places	&	H1	&	H1	&	H1	&	H0	&	H1	&	H0	&	H1	&	H1	&	H1	&	H0	&	H0	&	H0	\\
High-Literate-Women	&	H1	&	H1	&	H1	&	H0	&	H0	&	H0	&	H1	&	H1	&	H0	&	H0	&	H0	&	H0	\\
Main-Worker-Women	&	H1	&	H1	&	H0	&	H1	&	H1	&	H0	&	H1	&	H1	&	H1	&	H0	&	H1	&	H0	\\
		\hline
	\end{tabular}
	%}
	\label{TAB:Scaling_Compare_Classes}
\end{table}

%\newpage 
\bigskip
\noindent
{\bf{Author Contribution Statement: }} All authors contributed equally to the paper  \\
{\bf{Data Availability Statement: }} Data used here is publicly available, it is cited properly in the reference list. Data may be available on request.

%\vspace*{1cm}

\end{document}